\documentclass[usenatbib,onecolumn]{mn2e}

\usepackage{amssymb}
\usepackage{amsmath}
\usepackage{graphicx}
\usepackage{times}
\usepackage[colorlinks=true, linkcolor=blue, citecolor=red, urlcolor=blue]{hyperref}

\newcommand{\bfr}{{\boldsymbol{r}}}
\newcommand{\txd}{{\text{d}}}
\renewcommand{\leq}{\leqslant}

\begin{document}

\date{Accepted 2008 October 28. Received 2008 August 18}

\title{Exact potential-density pairs for flattened dark haloes}

\author[M.~Baes]{Maarten Baes \\ Sterrenkundig Observatorium,
Universiteit Gent, Krijgslaan 281-S9, B-9000 Gent, Belgium, maarten.baes@ugent.be}

\maketitle
\begin{abstract}
  Cosmological simulations suggest that dark matter haloes are not
  spherical, but typically moderately to strongly triaxial systems. We
  investigate methods to convert spherical potential-density pairs
  into axisymmetric ones, in which the basic characteristics of the
  density profile (such as the slope at small and large radii) are
  retained. We achieve this goal by replacing the spherical radius $r$
  by an oblate radius $m$ in the expression of the gravitational
  potential $\Phi(r)$. 

  We extend and formalize the approach pioneered by
  \citet{1975PASJ...27..533M} to be applicable to arbitrary
  potential-density pairs. Unfortunately, an asymptotic study
  demonstrates that, at large radii, such models always show a
  $R^{-3}$ disc superposed on a smooth roughly spherical density
  distribution. As a result, this recipe cannot be used to construct
  simple flattened potential-density pairs for dynamical systems such
  as dark matter haloes. Therefore we apply a modification of our
  original recipe that cures the problem of the discy behaviour. An
  asymptotic analysis now shows that the density distribution has the
  desired asymptotic behaviour at large radii (if the density falls
  less rapidly than $r^{-4}$). We also show that the flattening
  procedure does not alter the shape of the density distribution at
  small radii: while the inner density contours are flattened, the
  slope of the density profile is unaltered. 

  We apply this recipe to construct a set of flattened dark matter
  haloes based on the realistic spherical halo models by
  \citet{2005MNRAS.363.1057D}. This example illustrates that the
  method works fine for modest flattening values, whereas stronger
  flattening values lead to peanut-shaped density distributions.
\end{abstract}

\begin{keywords}
methods: analytical -- dark matter -- galaxies: haloes
\end{keywords}

\section{Introduction}

The intrinsic shape of dark matter haloes in the Cold Dark Matter
paradigm has been a matter of debate for the past two decades
\citep[e.g.][]{1988ApJ...327..507F, 1991ApJ...378..496D,
  1992ApJ...399..405W, 1996MNRAS.281..716C}. The most recent numerical
simulations of cosmological clustering in a $\Lambda$CDM framework
suggest that dark matter haloes are typically slowly rotating,
moderately to strongly triaxial systems, with an average flattening
that increases with increasing halo mass
\citep[e.g.][]{2002ApJ...574..538J, 2005ApJ...618....1H,
  2005ApJ...627..647B, 2005ApJ...629..781K, 2006MNRAS.367.1781A,
  2007MNRAS.377...50H}. In order to set up realistic models for dark
matter haloes, we need to leave the well-trodden path of spherical
symmetry and consider structures that reflect this variety in
shapes. A crucial step is the construction of exact flattened or
triaxial potential-density pairs with a realistic density structure
(for example a cuspy behaviour at small radii). The most
straightforward way to construct such axisymmetric or triaxial
potential-density pairs is to consider a spherical density profile
$\rho(r)$ with the desired characteristics. If one replaces in this
expression the spherical radius $r$ by an oblate spheroidal radius
\begin{equation}
  m(R,z)
  =
  \sqrt{R^2+\frac{z^2}{q^2}},
  \label{moblate}
\end{equation}
or a triaxial radius
\begin{equation}
  m(x,y,z)
  =
  \sqrt{x^2+\frac{y^2}{p^2}+\frac{z^2}{q^2}},
  \label{mtriaxial}
\end{equation}
one obtains an axisymmetric or triaxial density profile
$\rho_{\text{f}}(\bfr)\equiv\rho_{\text{f}}(m)$ whose isodensity
surfaces are concentric oblate spheroids or ellipsoids with fixed axis
ratios. This is a very straightforward recipe to construct simple
oblate, prolate or triaxial density profiles, but unfortunately the
formulae to calculate the corresponding potential
$\Phi_{\text{f}}(\bfr)$ are cumbersome and can only be calculated
numerically. In N-body or hydrodynamical simulations, the potential
(or its gradient) has to be calculated very frequently and should be
known with as efficiently and accurately as possible to avoid
numerical artifacts. As a result, it is worthwhile to look for methods
to create exact potential-density pairs for axisymmetric or triaxial
configurations in which the potential can be calculated in a more
straightforward way.

One general approach one can apply is to modify a spherical
potential-density pair by adding a (finite) spherical harmonics
expansion to it. While this approach can in principle be used to
construct exact potential-density pairs for axisymmetric or triaxial
systems \citep[e.g.][]{1996MNRAS.281.1333D}, the number of terms in
the expansion typically has to be substantial, such that we approach a
similar situation as before. An alternative approach to construct
axisymmetric or triaxial potential-density pairs is to apply a
transformation $r\rightarrow m$ on the potential rather than on the
density. If we have an analytical expression for
$\Phi_{\text{f}}(\bfr)\equiv\Phi_{\text{f}}(m)$, we can use Poisson's
equation to calculate the corresponding density distribution
$\rho_{\text{f}}(\bfr)$. The advantage of this approach is that the
calculation of the density $\rho_{\text{f}}(\bfr)$ from the potential
$\Psi_{\text{f}}(m)$ requires only differentiations, such that one
always obtains potential-density pairs in which both potential and
density are analytical functions. One disadvantage is that one cannot
a priori set the shape of the isodensity contours; it is not even
guaranteed that the density corresponding to a given potential is
everywhere positive. A second disadvantage is that simple recipes such
as equation~(\ref{moblate}) do not work for the potential, because the
isopotential surfaces need to become spherical at large radii. Other,
unavoidably more complicated, recipes must be considered.

A successful example to construct an oblate analytical
potential-density pair based on this strategy was the seminal work by
\citet{1975PASJ...27..533M}. These authors create an exact
potential-density pair by considering the spherical Plummer potential
\begin{equation}
  \Phi(r)
  =
  -\frac{GM}{\sqrt{r^2+b^2}},
\end{equation}
and turning it into a flattened potential
\begin{equation}
  \Phi_{\text{f}}(R,z)
  =
  -\frac{GM}{\sqrt{R^2+\left(a+\sqrt{z^2+b^2}\right)^2}}.
\end{equation}
The resulting density distribution can be calculated fairly easily.
This technique of constructing flattened or triaxial potential-density
pairs has been applied to other potentials by e.g.\
\citet{1980PASJ...32...41S}, \citet{2000MNRAS.319.1067Z} and
\citet{2007PASJ...59..319V}. While this recipe provides a
straightforward means to create analytical flattened potential-density
pairs, it generates densities that do not correspond to astrophysical
systems such as dark matter haloes. Indeed, at large radii the
Miyamoto-Nagai type models show a notorious double behaviour, with a
thin disc superposed on a roughly spherical density profile. While
this could provide a first approximation for disc galaxies (although
the density in galaxy discs is exponential rather than $R^{-3}$), this
is obviously not the desired shape for a dark matter halo.

The goal of the present paper is to construct axisymmetric or triaxial
potential-density pairs that can be used to model realistic dark
matter haloes, starting from an arbitrary spherical potential-density
pair. In Section~2 we formulate a recipe based on the
\citet{1975PASJ...27..533M} approach and we demonstrate that such
models always have the conspicuous $R^{-3}$ disc-like behaviour at
large radii, such that it cannot be used to represent dynamical
systems as dark matter haloes. In Section~3 we adapt our recipe to
cure this conspicuous behaviour and we demonstrate that this new
recipe can produce analytical axisymmetric potential-density pairs
which retain the original behaviour of the original density profile at
both small and large radii. In Section~4 we apply this recipe to
create a set of flattened potential-density suitable to represent dark
matter haloes based on the spherical \citet{2005MNRAS.363.1057D}
models. Section~5 sums up.

\section{Miyamoto-Nagai type models}
\label{MN.sec}

In order to create a set of oblate spheroidal models, we will replace
the spherical radius $r$ in the spherical model by a spheroidal radius
$m=m(R,z)$. We need to find a prescript such that $m$ is flattened at
small radii and spherical at large radii. A general form that
satisfies these requirements is
\begin{equation}
  m
  \equiv
  m(R,z)
  =
  \sqrt{R^2+\frac{z^2}{Q^2(z)}},
  \label{mgeneral}
\end{equation}
where $Q(z)$ is a monotonically rising function of $z$ which has the
asymptotic behaviour that it approaches 1 at large radii and a
constant value $q$ at small radii. There are of course infinitely
many different functions $Q(z)$ with these characteristics.
Inspired by \citet{1975PASJ...27..533M} we concentrate on one such
particular function
\begin{equation}
  Q(z)
  =
  \frac{z}{\sqrt{\left(a^2-b^2+\sqrt{z^2+b^2}\right)^2-a^2}},
  \label{Q}
\end{equation}
for which we find
\begin{equation}
  q
  =
  \sqrt{\frac{b}{a}}.
  \label{q}
\end{equation}
Substituting expression~(\ref{Q}) into (\ref{mgeneral}) we obtain
\begin{equation}
  m(R,z)
  =
  \sqrt{R^2+\left(a-b+\sqrt{z^2+b^2}\right)^2-a^2}.
  \label{m}
\end{equation}
When we move to polar coordinates in the equatorial plane using
$R=r\sin\theta$ and $z=r\cos\theta$, we find that at small radii
\begin{equation}
  m(r,\theta) 
  \sim
  \frac{\sqrt{\cos^2\theta+q^2\sin^2\theta}}{q}\,r,
  \label{msmallr}
\end{equation}
whereas at large radii $m(r,\theta)\sim r$. The equipotential surfaces
of the potential $\Phi_{\text{f}}(r,\theta)=\Phi_{\text{f}}(m)$ will
hence be flattened with flattening $q$ at small radii and spherical at
large radii, as required.

Once we have constructed the potential $\Phi_{\text{f}}(r,\theta)$, we
can compute the corresponding density $\rho_{\text{f}}(r,\theta)$ that
self-consistently generates this potential through Poisson's equation.
\begin{equation}
  \rho_{\text{f}}(r,\theta)
  =
  \frac{1}{4\pi r^2}\left[
    \frac{\partial}{\partial r}
    \left(
      r^2\,
      \frac{\partial m}{\partial r}\,
      \frac{\txd\Phi_{\text{f}}}{\txd m}
    \right)
    +
    \frac{1}{\sin^2\theta}\,\frac{\partial}{\partial\theta}
    \left(
      \sin\theta\,
      \frac{\partial m}{\partial\theta}\,
      \frac{\txd\Phi_{\text{f}}}{\txd m}
    \right)
  \right].
\end{equation}
After some manipulation we can write this equation as
\begin{equation}
  \rho_{\text{f}}(r,\theta)
  =
  T_1(r,\theta)\,\rho(m)
  +
  T_2(r,\theta)\,\bar\rho(m).
  \label{rhof}
\end{equation}
Here, $\rho(m)$ represents the density of the original spherical model
evaluated at $r=m$ and $\bar\rho(m)$ is the average density of the
original spherical model within a radius $r$,
\begin{equation}
  \bar\rho(r)
  =
  \frac{M(r)}{\frac{4\pi}{3}r^3},
  \label{barrho}
\end{equation}
also evaluated at $r=m$. The coefficients $T_1$ and $T_2$ in
equation~(\ref{rhof}) are independent of the potential of the system,
and are given by
\begin{equation}
  T_1(r,\theta)
  =
  \frac{r^2}{m^2}
  \left[\sin^2\theta
    +\left(1-\frac{G}{2F}\right)^2\frac{\cos^2\theta}{F^2}\right],
  \label{T1}
\end{equation}
and
\begin{equation}
  T_2(r,\theta)
  =
  \frac{r^2}{3m^2}
  \left[
    \left(
      \frac{1}{F}-1-\frac{2G}{F^2}-\frac{H}{2F^2}+\frac{G^2}{F^3}
    \right)
    \sin^2\theta
    +
    \left(
      2-\frac{2}{F}+\frac{G}{F^2}-\frac{H}{2F^2}+\frac{G^2}{4F^3}
    \right)
    \frac{\cos^2\theta}{F}
  \right],
  \label{T2}
\end{equation}
with 
\begin{gather}
  F=Q^2(z), 
  \\
  G=z\,F'(z),
  \\
  H=z^2\,F''(z).
\end{gather}
These coefficients are never divergent and it is straightforward to
check that for the trivial case $a=b$, they reduce to $T_1\equiv1$ and
$T_2\equiv0$, such that we recover the original spherical density
distribution $\rho(r)$, as required.

The asymptotic expansion for the two coefficients $T_1$ and $T_2$ at
large radii reads
\begin{gather}
  T_1(r,\theta)
  \sim
  1
  +
  \frac{(a-b)\,|\cos\theta|}{r},
  \label{T1larger}
  \\
  T_2(r,\theta)
  \sim
  -\frac{a^2-b^2}{r^2}.
  \label{T2larger}
\end{gather}
If the initial spherical model has a finite mass, the mean density
$\bar\rho(m)$ behaves asymptotically as $r^{-3}$. If the density of
the initial spherical model decreases more slowly than $r^{-5}$, the
density of the corresponding oblate model will asymptotically reduce
to the density of the spherical model.

There is one caveat, however: the asymptotic expansion for the
coefficients $T_1$ and $T_2$ derived above is not valid within the
equatorial plane $\theta=\tfrac{\pi}{2}$. Since $z=0$ in the entire
equatorial plane, we have by construction $F=q^2$ and $G=H=0$ at every
position $R=r=m$ in the equatorial plane, such that we have the
identities
\begin{gather}
    T_1\left(R,\tfrac{\pi}{2}\right)
    \equiv
    1,
    \\
    T_2\left(R,\tfrac{\pi}{2}\right)
    \equiv
    \frac{1-q^2}{3q^2}
    =
    \frac{a-b}{3b}.
\end{gather}
In the equatorial plane, the contribution of the second term in
(\ref{rhof}) will not decrease as $R^{-5}$, but only as $R^{-3}$, and
it will therefore dominate the contribution of the first term. The
mean density term $\bar\rho(m)$ is only dominant at large radii in the
equatorial plane, such that the final density
$\rho_{\text{f}}(r,\theta)$ will have a discy structure at large
radii, with a thin $R^{-3}$ disc superposed on a nearly spherical
density profile. This behaviour is present in all Miyamoto-Nagai type
models \citep{1975PASJ...27..533M, 1980PASJ...32...41S,
  2000MNRAS.319.1067Z} and unfortunately turns this recipe unsuitable
for the construction of realistic flattened dark matter halo models.

\section{A new recipe to construct flattened models}
\label{new.sec}

\subsection{Construction}

An inconvenience about the family of models presented in the previous
section was the appearance of a discy structure at large radii. At
large radii, the density profiles of the models behave similarly as
their spherical progenitors, at all polar angles except in the
equatorial plane $z=0$. Mathematically, the reason for this discy
structure is the discontinuous asymptotic behaviour of the function
$Q(z)$. Outside the equatorial plane, $|z|\rightarrow\infty$ and
$Q\rightarrow1$ at large radii, such that the coefficient $T_2$
disappears. In the equatorial plane, however, $z$ is always equal to
zero and $T_2$ does not converge to zero, such that the mean density
term $\bar\rho(m)$ contributes to (and dominates) the density.

If we want to construct a set of models which do not show this discy
structure at large radii, we can try to adapt the recipe from the
previous section in a way that $Q$ smoothly goes to one at large
radii, both inside and outside the equatorial plane. This can be
achieved by replacing the oblate radius $m$ from
equation~(\ref{mgeneral}) by
\begin{equation}
  m(R,z)
  =
  \sqrt{R^2+\frac{z^2}{Q^2(r)}}.
  \label{mgeneral2}
\end{equation}
Using the functional form (\ref{Q}) for $Q(r)$, we obtain
\begin{equation}
  m(R,z)
  =
  \sqrt{R^2+\frac{z^2}{r^2}
    \left[\left(a-b+\sqrt{r^2+b^2}\right)^2-a^2\right]}.
  \label{m2}
\end{equation}
or
\begin{equation}
  m(r,\theta)
  =
  \sqrt{r^2\sin^2\theta+\cos^2\theta
    \left[\left(a-b+\sqrt{r^2+b^2}\right)^2-a^2\right]}.
  \label{m2}
\end{equation}
This new oblate radius $m$ has a similar asymptotic behaviour at small
and large radii as the original version (\ref{m}). Replacing $r$ by
this oblate radius $m$ in the expression $\Phi(r)$ for the potential
of a spherical model, we obtain an oblate axisymmetric potential
$\Phi_{\text{f}}(r,\theta)$. The density that self-consistently
generates this potential can be found through Poisson's
equation. Going through a similar analysis as in the previous Section,
we find that the density can be written as
\begin{equation}
  \rho_{\text{f}}(r,\theta)
  =
  U_1(r,\theta)\,\rho(m)
  +
  U_2(r,\theta)\,\bar\rho(m),
  \label{rhof2}
\end{equation}
where the coefficients $U_1$ and $U_2$ are now given by
\begin{equation}
  U_1(r,\theta)
  =
  \frac{r^2}{m^2}
  \left[
    \sin^2\theta
    +
    \frac{1-G}{F^2}\,\cos^2\theta
    +
    \frac{G}{F^2}\left(1-\frac{1}{F}+\frac{G}{4F^2}\right)\cos^4\theta
  \right],
  \label{U1}
\end{equation}
and
\begin{equation}
  U_2(r,\theta)
  =
  \frac{r^2}{3m^2}
  \left[
    \left(\frac{1}{F}-1\right)
    +
    \left(1+\frac{1}{F}-\frac{2}{F^2}-\frac{H}{2F^2}+\frac{G^2}{F^3}\right)
    \cos^2\theta
    +
    \frac{1}{F^2}
    \left(\frac{H}{2}-\frac{G^2}{F}-\frac{H}{2F}+\frac{G^2}{4F^2}\right)
    \cos^4\theta
  \right],
  \label{U2}
\end{equation}
with now obviously
\begin{gather}
  F=Q^2(r), 
  \\
  G = r\,F'(r),
  \\
  H=r^2\,F''(r).
\end{gather}
Again, it is elementary to check that for $U_1\equiv1$ and
$U_2\equiv0$ for the trivial case $a=b$.

\subsection{The asymptotic behaviour at large radii}

After some algebra, one obtains that the asymptotic expansion of the
coefficients $U_1$ and $U_2$ at large radii becomes
\begin{gather}
  U_1(r,\theta)
  \sim
  1,
  \label{U1larger}
  \\
  U_2(r,\theta)
  \sim
  -\frac{2\,(a-b)\cos2\theta}{3r}.
  \label{U2larger}
\end{gather}
These expansions are valid for all directions; in particular, there is
no different behaviour of these components in the equatorial
plane. Comparing the asymptotic behaviour of these coefficients to
those of the previous Section, we see that the coefficients $T_1$ and
$U_1$ both have an asymptotic behaviour with simply 1 as the leading
term, such that the first term in the total density~(\ref{rhof})
or~(\ref{rhof2}) behaves identically as the density of the original
spherical model. The coefficients $T_2$ and $U_2$ behave in a
different way, however: while $T_2$ decreases as $r^{-2}$ at large
radii, $U_2$ decreases only as $r^{-1}$, with an additional
$\cos2\theta$ azimuthal dependence. Only for those models where the
density of the initial spherical model decreases more slowly than
$r^{-4}$, the density of the corresponding flattened model will
asymptotically reduce to the density of the spherical model. For
models where the density decreases more rapidly at large radii, the
density will be dominated by the second term. As this term becomes
negative at large radii near the rotation axis, such models are
unacceptable. This method to flatten spherical potential-density pairs
is therefore limited to those models where the density decreases more
slowly than $r^{-4}$. This is typically the case for dark matter
haloes.

\subsection{The asymptotic behaviour at small radii}

The goal of this paper was to consider a method to, starting from a
spherical potential-density pair, construct a flattened analog that
preserves the asymptotic behaviour of the density profile at both
small and large radii. An asymptotic analysis of the functions $U_1$
and $U_2$ learns that both terms converge to a finite value at small
radii,
\begin{gather}
  U_1(r,\theta)
  \sim
  \frac{1}{q^2}\,
  \frac{\cos^2\theta + q^4\sin^2\theta}
  {\cos^2\theta+q^2\sin^2\theta},
  \label{U1smallr}
  \\
  U_2(r,\theta)
  \sim
  \frac{1-q^2}{3q^2}\,
  \frac{q^2\sin^2\theta - 2\cos^2\theta}
  {\cos^2\theta+q^2\sin^2\theta}.
  \label{U2smallr}
\end{gather}
Assume that our density profile $\rho(r)$ behaves as a power law at
small radii,
\begin{equation}
  \rho(r)
  \sim
  \rho_0r^{-\gamma_0},
  \label{rhosmallr}
\end{equation}
with $0\leq\gamma_0<3$. It is easy to check that the mean density
$\bar\rho(r)$ has a similar slope,
\begin{equation}
  \bar\rho(r)
  \sim
  \frac{3\rho_0}{3-\gamma_0}\,r^{-\gamma_0}.
  \label{barrhosmallr}
\end{equation}
Combining the expressions~(\ref{msmallr}), (\ref{rhof2}),
(\ref{U1smallr}), (\ref{U2smallr}), (\ref{rhosmallr}) and
(\ref{barrhosmallr}), we find the asymptotic behaviour of
$\rho_{\text{f}}(r,\theta)$ at small radii,
\begin{equation}
  \rho_{\text{f}}(r,\theta)
  \sim
  \frac{\rho_0}{3-\gamma_0}
  \frac{(1-\gamma_0 q^2+2q^2)\,q^2\sin^2\theta+(1-\gamma_0+2q^2)\cos^2\theta}
  {q^{2-\gamma_0}\,(\cos^2\theta+q^2\sin^2\theta)^{1+\gamma_0/2}}\,
  r^{-\gamma_0},
\end{equation}
This result demonstrates that the asymptotic behaviour of the original
spherical model is preserved after the flattening: the slope of the
density profile remains, only the zero point changes and becomes
dependent of the polar angle $\theta$. In particular, along the major
and minor axes we find the expansions
\begin{gather}
  \rho_{\text{f}}(R,\tfrac{\pi}{2})
  \sim
  \rho_0\,
  \frac{1-\gamma_0 q^2+2q^2}{(3-\gamma_0)\,q^2}\,
  R^{-\gamma_0},
  \\
  \rho_{\text{f}}(z,0)
  \sim
  \rho_0\,
  \frac{1-\gamma_0+2q^2}{(3-\gamma_0)\,q^{2-\gamma_0}}\,
  z^{-\gamma_0},
\end{gather}
Using these expressions, we easily find the expression for the
flattening $q_\rho$ of the isodensity surfaces at small radii,
\begin{equation}
  q_\rho
  \sim
  \left(\frac{1-\gamma_0+2q^2}{1-\gamma_0 q^2+2q^2}\right)^{1/\gamma_0} q.
\end{equation}
For modest flattening ($q\lesssim1$) we find a nearly linear relation,
\begin{equation}
  1-q_\rho
  \sim
  \left(\frac{5-\gamma_0}{3-\gamma_0}\right) (1-q).
\end{equation}
Apart from the well-known fact
that the isodensity surfaces are generally more flattened than the
isopotential surfaces, this 
equation illustrates that the flattening of the inner isodensity
surfaces increases with increasing $\gamma_0$.


\section{Application: flattened Dehnen-McLaughlin dark halo models}

As an application of the formalism we have described in the previous
section, we construct in this section a set of flattened dark matter
halo models. Our starting point is the general three-parameter family
of generalized NFW models or Zhao models
\begin{equation}
  \rho(r)
  =
  \rho_0\,\left(\frac{r}{r_0}\right)^{-\gamma_0}
  \left[
    1+\left(\frac{r}{r_0}\right)^{\eta}
  \right]^{-(\gamma_\infty-\gamma_0)/\eta}.
  \label{zhao}
\end{equation}
The density of this general model is characterized by a power law with
(negative) slopes $\gamma_0$ and $\gamma_\infty$ at respectively small
and large radii. The third parameter, $\delta$, is a measure of the
width of the transition region between the two power-law zones. This
three-parameter model, the general dynamical properties of which are
described by \citet{1993MNRAS.262.1062S} and
\citet{1996MNRAS.278..488Z}, includes a large variety of well-known
simple potential density pairs used extensively throughout the
literature to describe dynamical systems ranging from galactic nuclei
to dark matter haloes \citep[e.g.][]{1911MNRAS..71..460P,
  1983MNRAS.202..995J, 1990ApJ...356..359H, 1993MNRAS.265..250D,
  1994AJ....107..634T, 1997ApJ...490..493N}. 

To demonstrate our recipe to construct oblate dark matter haloes, we
focus on a particular, non-trivial subset of this general
three-parameter set of models presented by
\citet{2005MNRAS.363.1057D}. These authors used the Jeans equations to
derived the density profile of a self-gravitating spherically
symmetric dynamical system that satisfy two characteristics of dark
haloes observed in cosmological simulations. The first of these
characteristics is the mysterious power-law behaviour of the so-called
pseudo phase-space density\footnote{In fact,
  \citet{2005MNRAS.363.1057D} assumed a power-law behaviour of the
  quantity $\rho/\sigma_r^\epsilon$ with $\sigma_r$ the radial
  velocity dispersion and $\epsilon$ a free parameter. We assume
  $\epsilon=3$, the most natural choice.} $Q = \rho/\sigma^3$ with
$\sigma(r)$ the velocity dispersion \citep{2001ApJ...563..483T,
  2004MNRAS.352.1109A, 2004MNRAS.351..237R}. The second ingredient in
their model is a linear relation between the logarithmic density slope
and the anisotropy, observed in different haloes with radically
different formation histories \citep{2006NewA...11..333H,
  2006JCAP...05..014H}. Constrained by these two assumptions, together
with the assumption of spherical symmetry, \citet{2005MNRAS.363.1057D}
derived a dark matter halo density of the form~(\ref{zhao}), with
\begin{gather}
  \gamma_0 = \frac{7+10\beta_0}{9},
  \\
  \gamma_\infty = \frac{31-20\beta_0}{9},
  \\
  \eta = \frac{4-2\beta_0}{9},
  \\
  \rho_0 = \frac{10-5\beta_0}{18\pi}\,\frac{M}{r_0^3}
\end{gather}
with $\beta_0$ the value of the velocity anisotropy parameter at small
radii and $M$ the total mass. The gravitational potential that
generates this mass density is
\begin{equation}
  \Phi(r) 
  =
  -\frac{9}{4-2\beta_0}\,\frac{GM}{r_0}\,
  B_{\frac{1}{1+(r/r_0)^\eta}}
  \left(\frac{9}{4-2\beta_0},\frac{11-10\beta_0}{4-2\beta_0}\right),
\end{equation}
with $B_x(p,q)$ the incomplete Beta function. 

\begin{figure*}
  \includegraphics[width=0.245\textwidth]{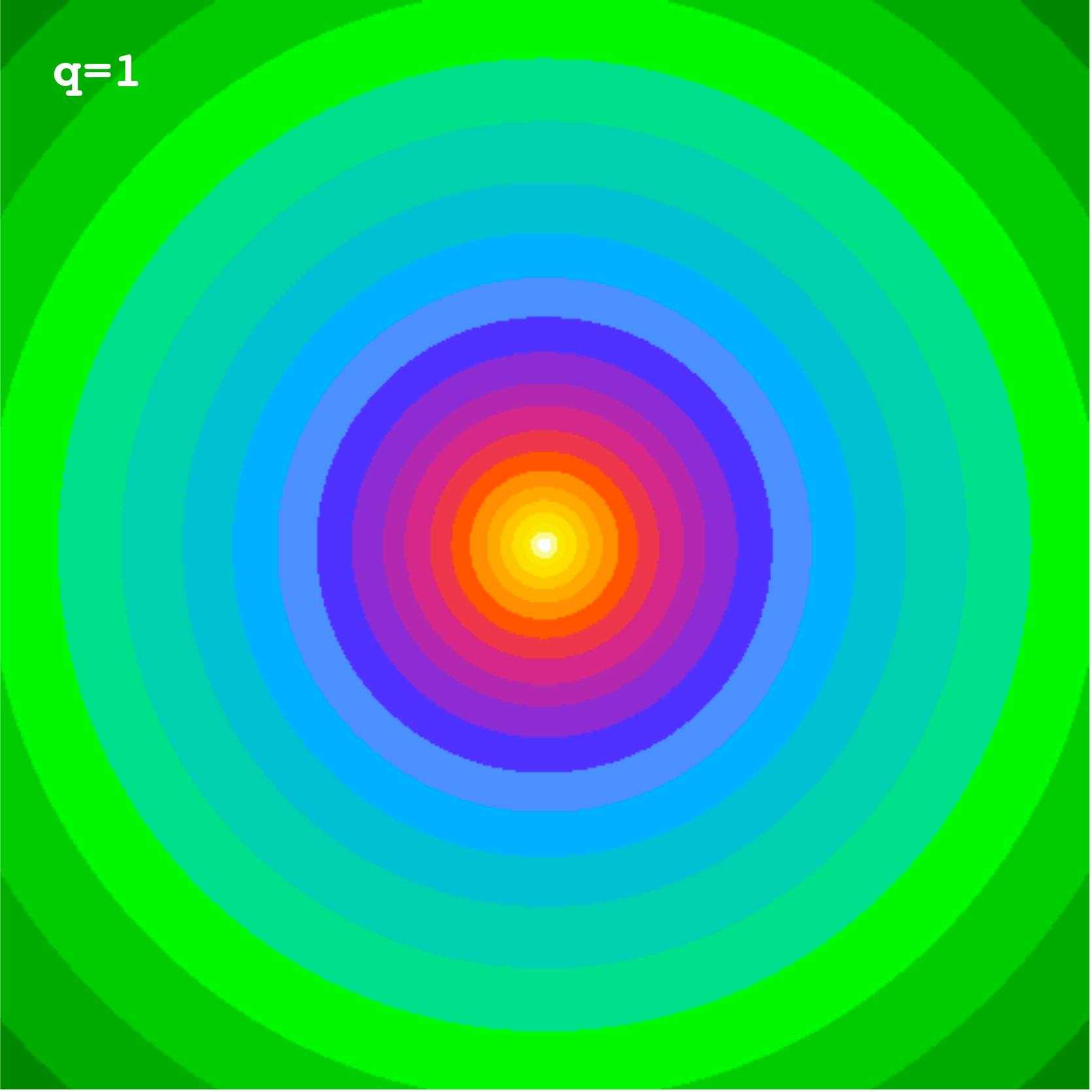}
  \includegraphics[width=0.245\textwidth]{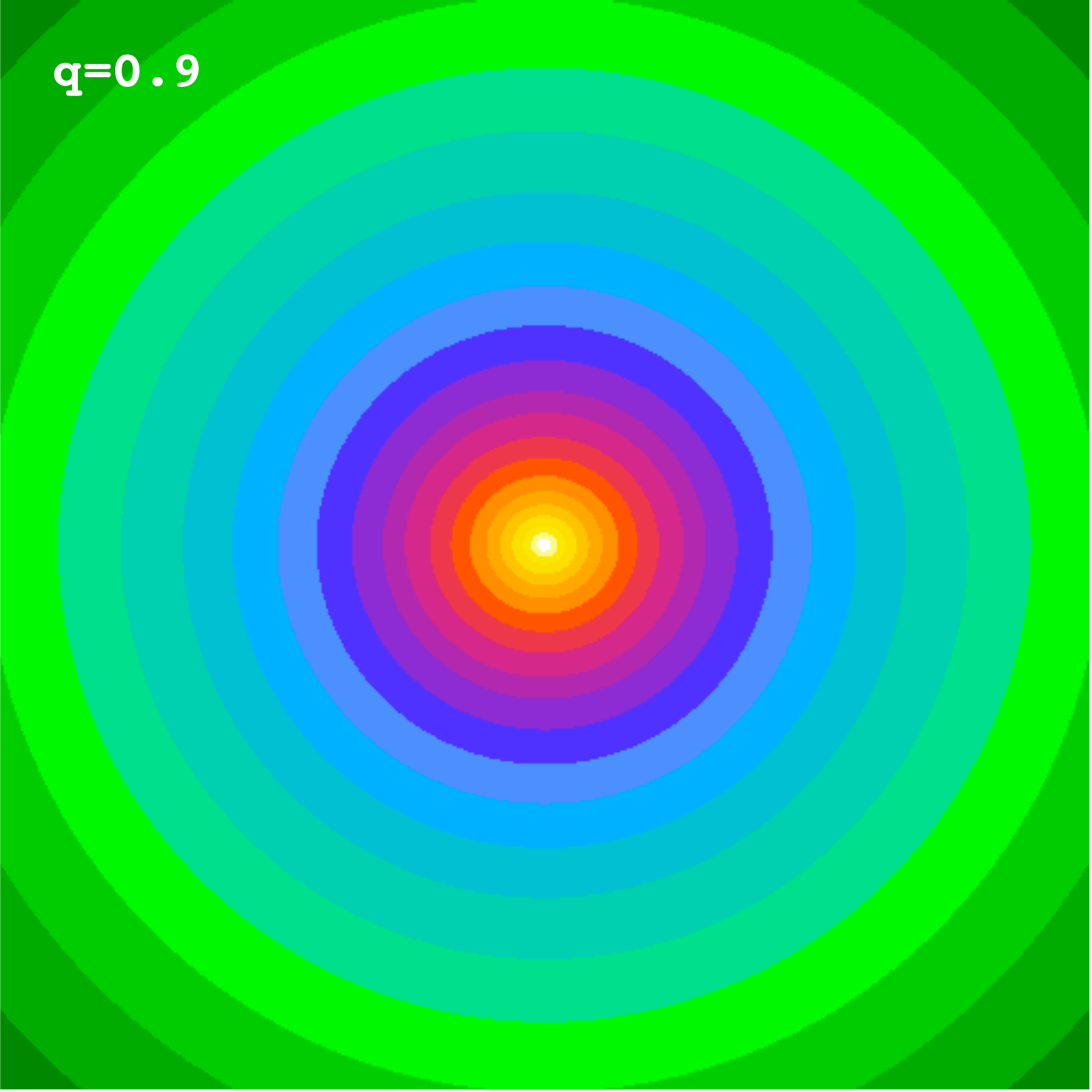}
  \includegraphics[width=0.245\textwidth]{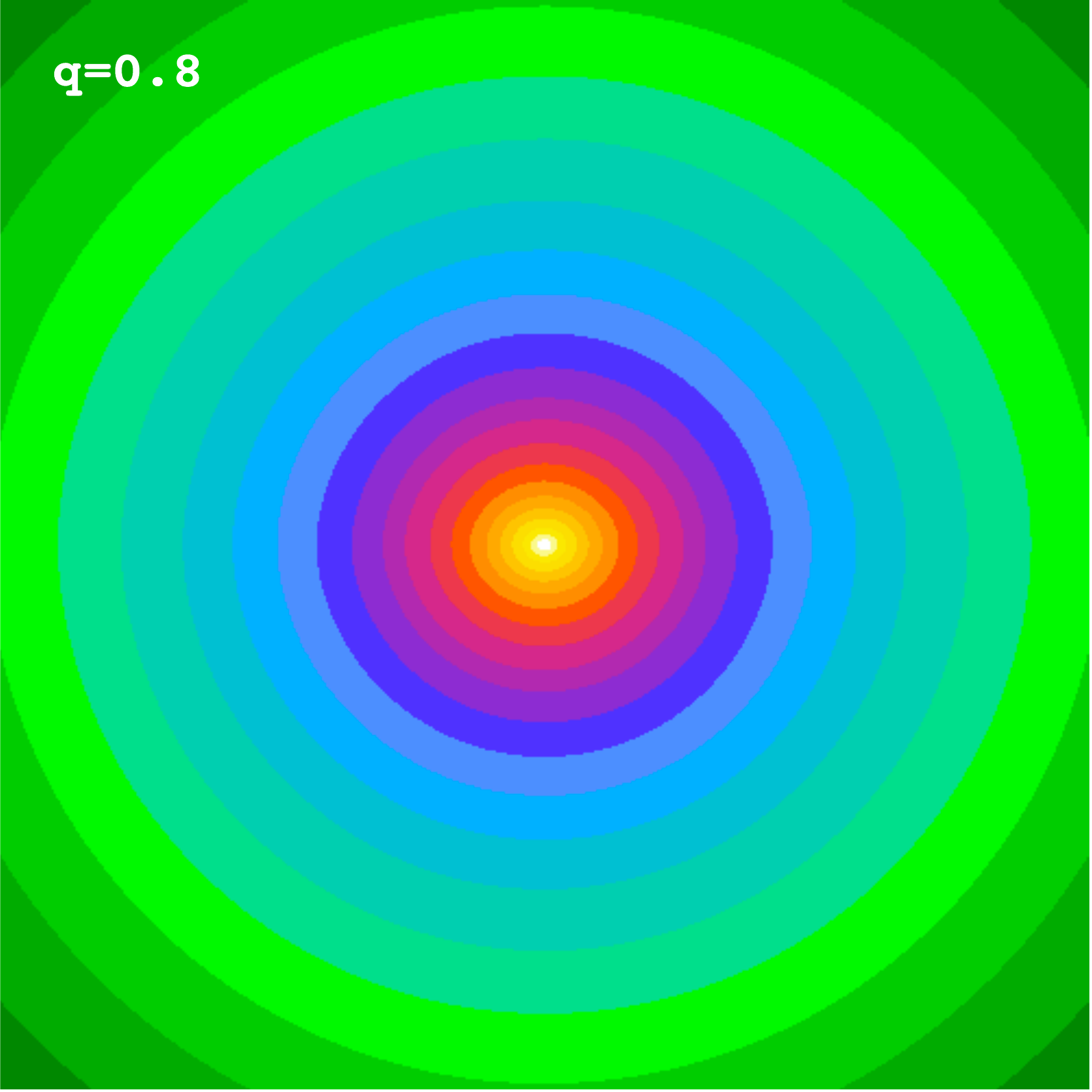}
  \includegraphics[width=0.245\textwidth]{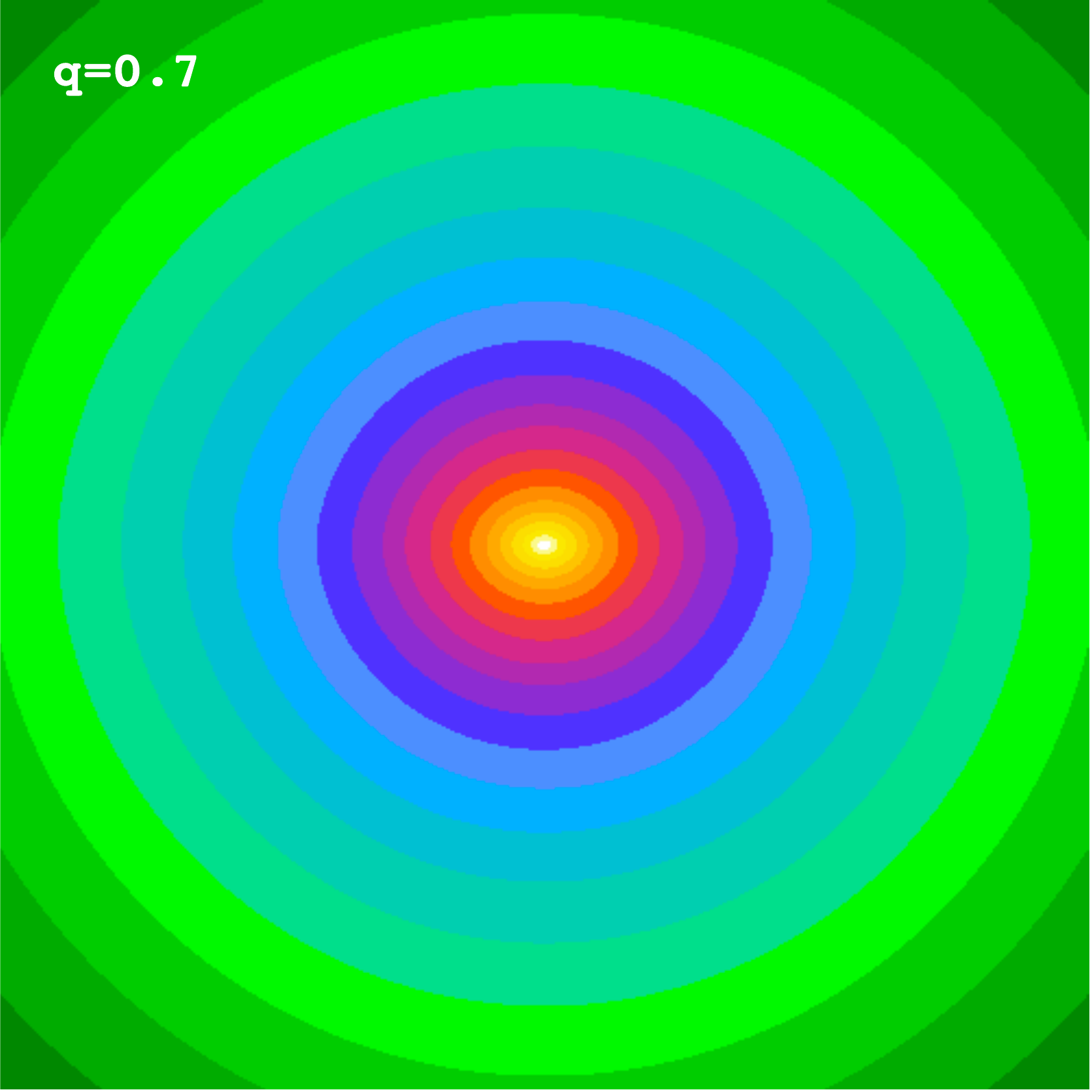}
  \\[0.2mm]
  \includegraphics[width=0.245\textwidth]{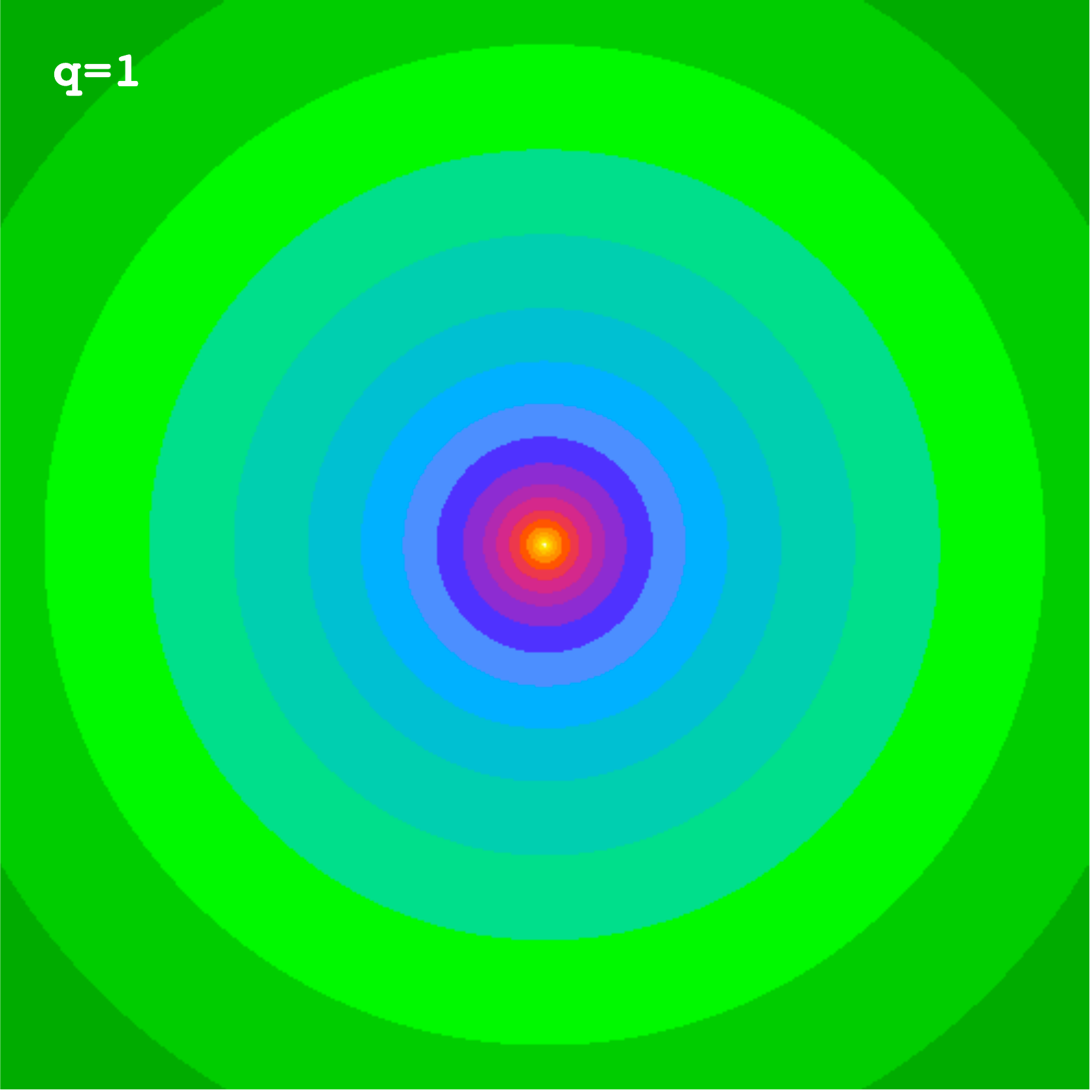}
  \includegraphics[width=0.245\textwidth]{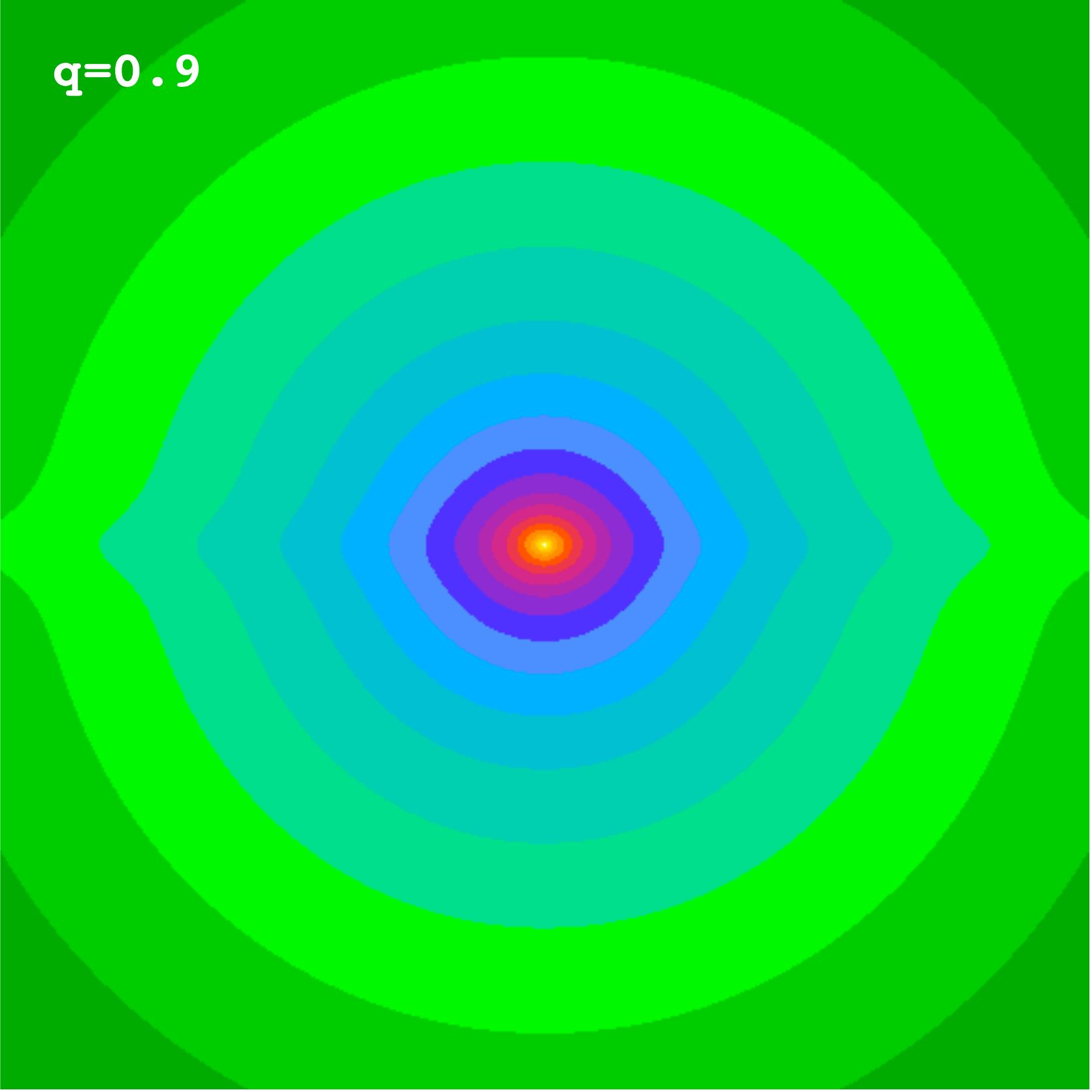}
  \includegraphics[width=0.245\textwidth]{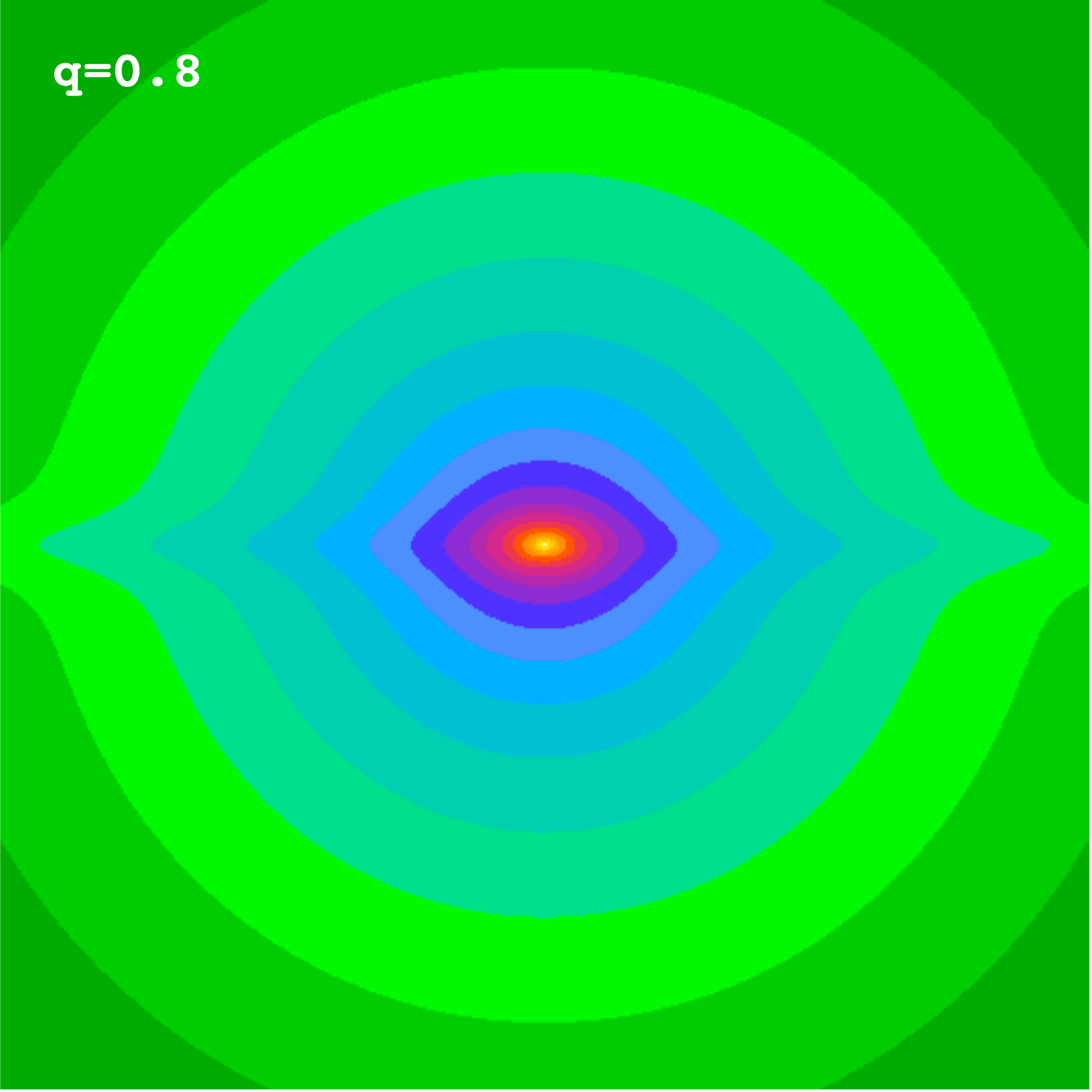}
  \includegraphics[width=0.245\textwidth]{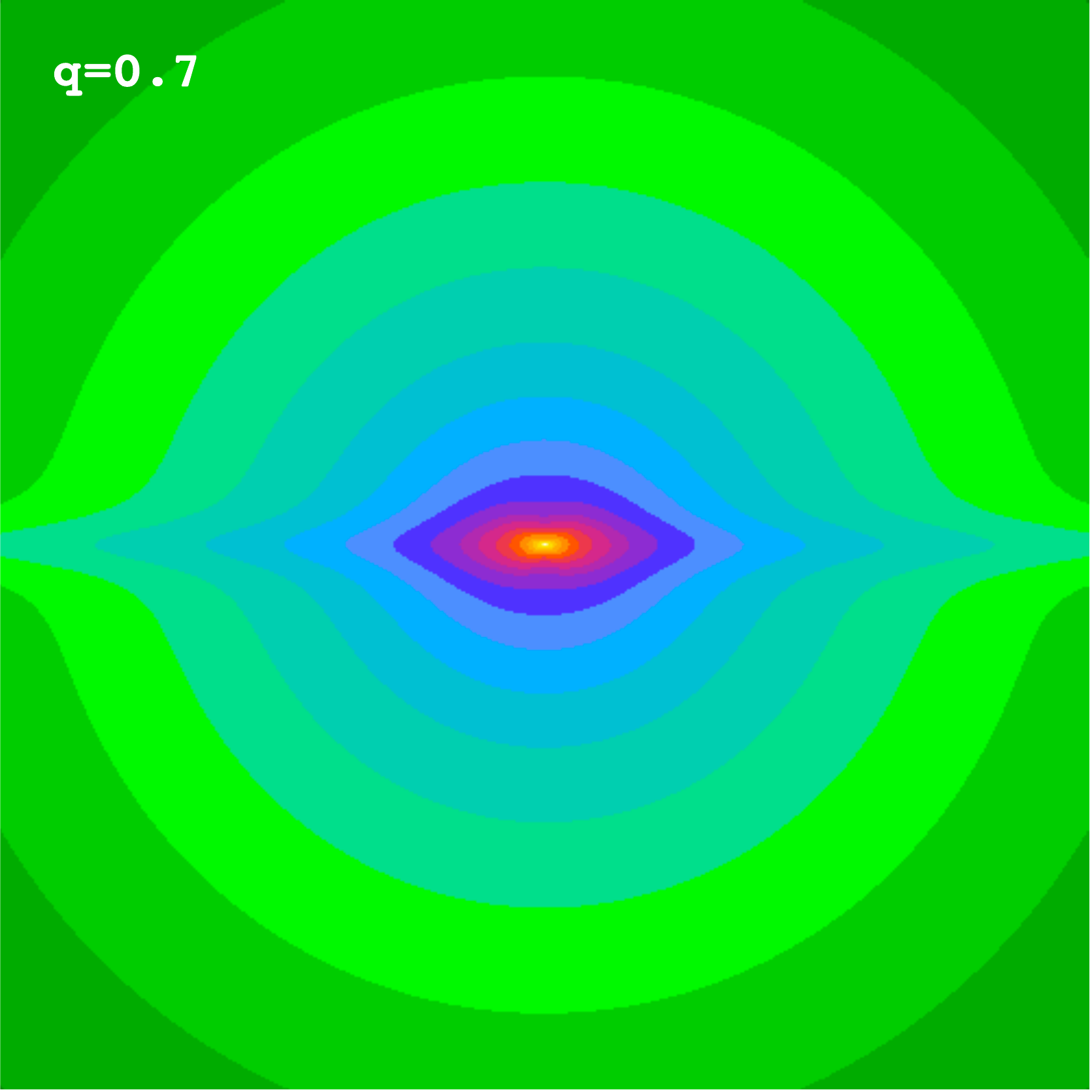}
  \caption{Isopotential (top row) and isodensity (bottom row) plots in
    the meriodional plane of the \citet{2005MNRAS.363.1057D} halo
    models flattened according to the Miyamoto-Nagai type flattening
    from Section~{\ref{MN.sec}}. Different models are shown according
    to different values of the potential flattening parameter $q$,
    ranging from 1 to 0.7. Each plot has dimensions of
    $10a\times10a$. At large radii, the isodensity surfaces are discy
    with an $R^{-3}$ dependence in the equatorial plane.}
  \label{potdensMN.pdf}
\end{figure*}

We constructed flattened versions of the spherical
\citet{2005MNRAS.363.1057D} haloes with $\beta_0=0$ and $r_0=a$ using
the recipes of Sections~2 and 3. In Figure~{\ref{potdensMN.pdf}} we
plot the isopotential and isodensity surfaces in the meridional plane
for the Miyamoto-Nagai type models for different values of the
parameter $q=b/a$. The isopotential surfaces (top row) are nearly
spheroidal, and the flattening decreases smoothly from the centre
(where the axis ratio is equal to $q$) to spherical at large
radii. The bottom row shows the isodensity surface plots of the
corresponding models. The isodensity surfaces are nearly spheroidal in
the central regions with a flattening that is much stronger than the
flattening of the potential. For increasing radii, the isodensity
surfaces become increasingly lemon-shaped, and at large radii, they
become very discy with an $R^{-3}$ dependence in the equatorial plane,
in agreement with the results from Section~2. Obviously, these models
can not represent realistic dark haloes.

\begin{figure*}
  \includegraphics[width=0.245\textwidth]{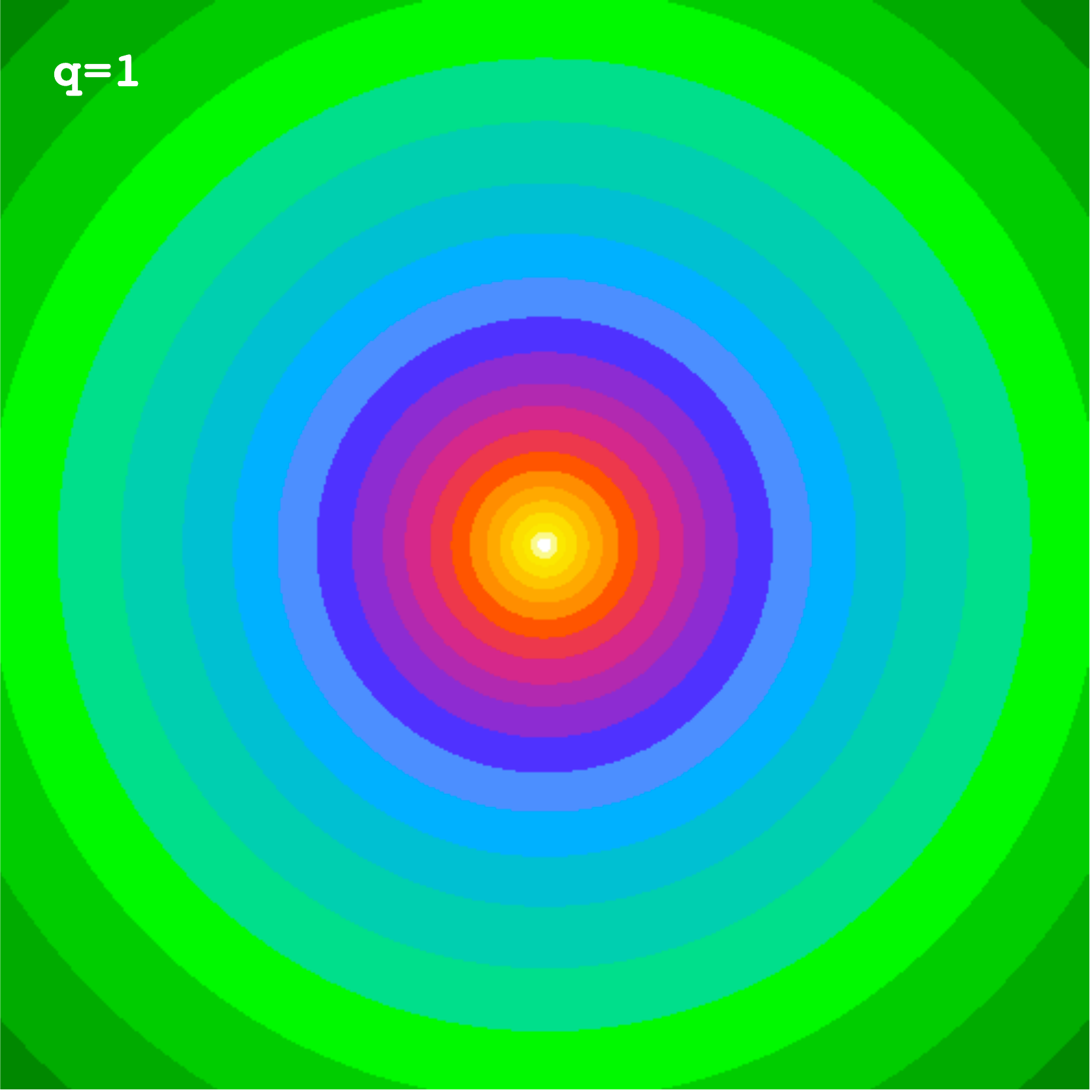}
  \includegraphics[width=0.245\textwidth]{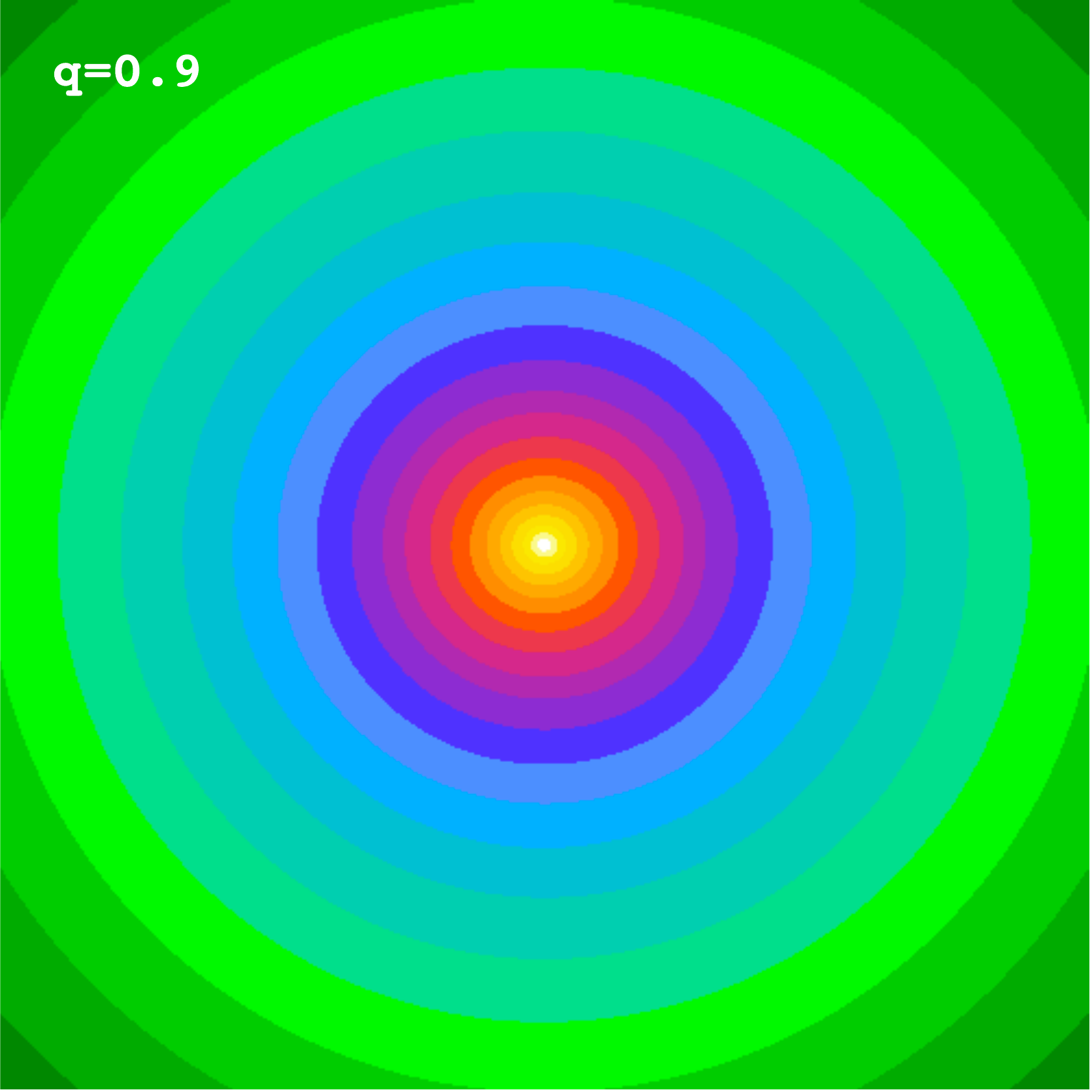}
  \includegraphics[width=0.245\textwidth]{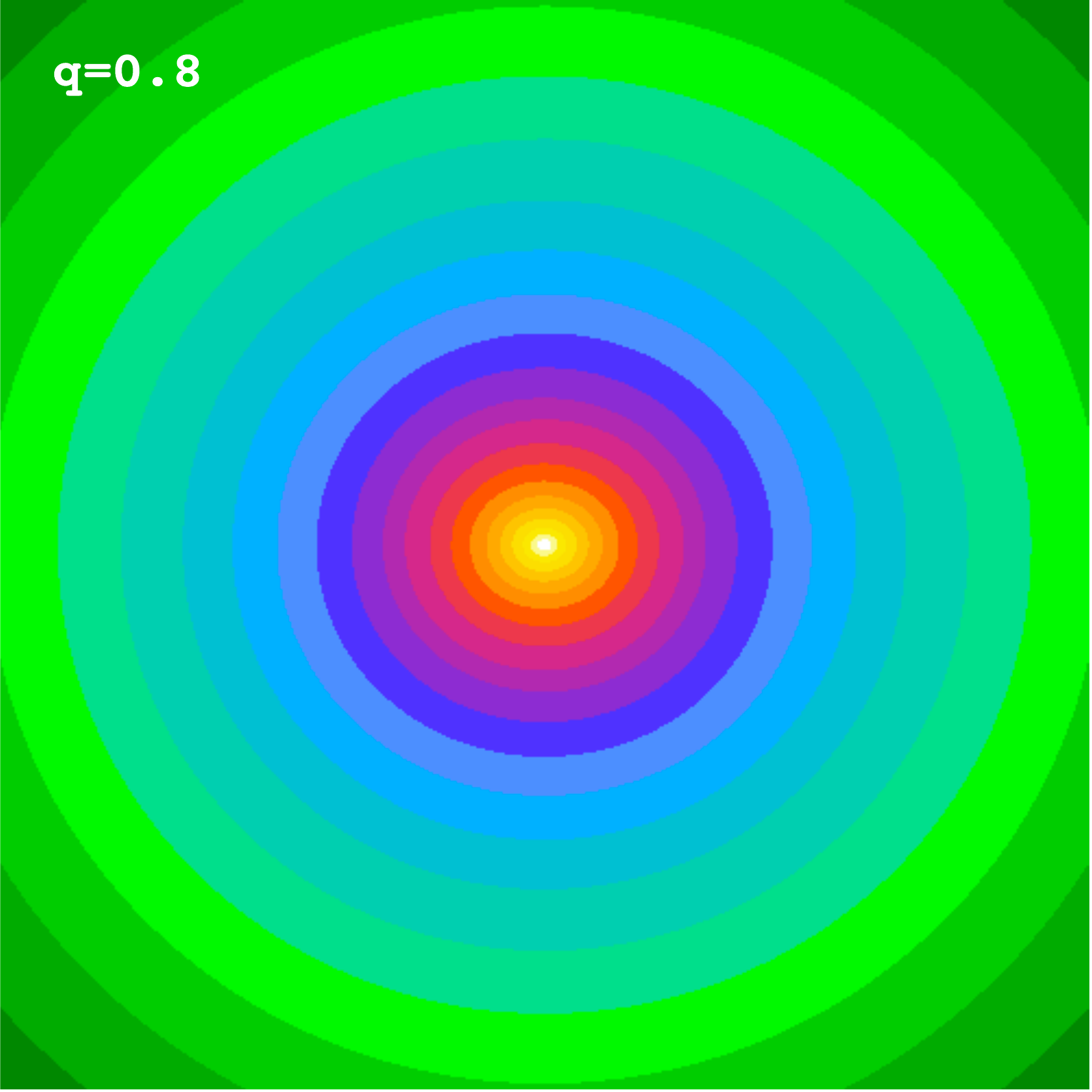}
  \includegraphics[width=0.245\textwidth]{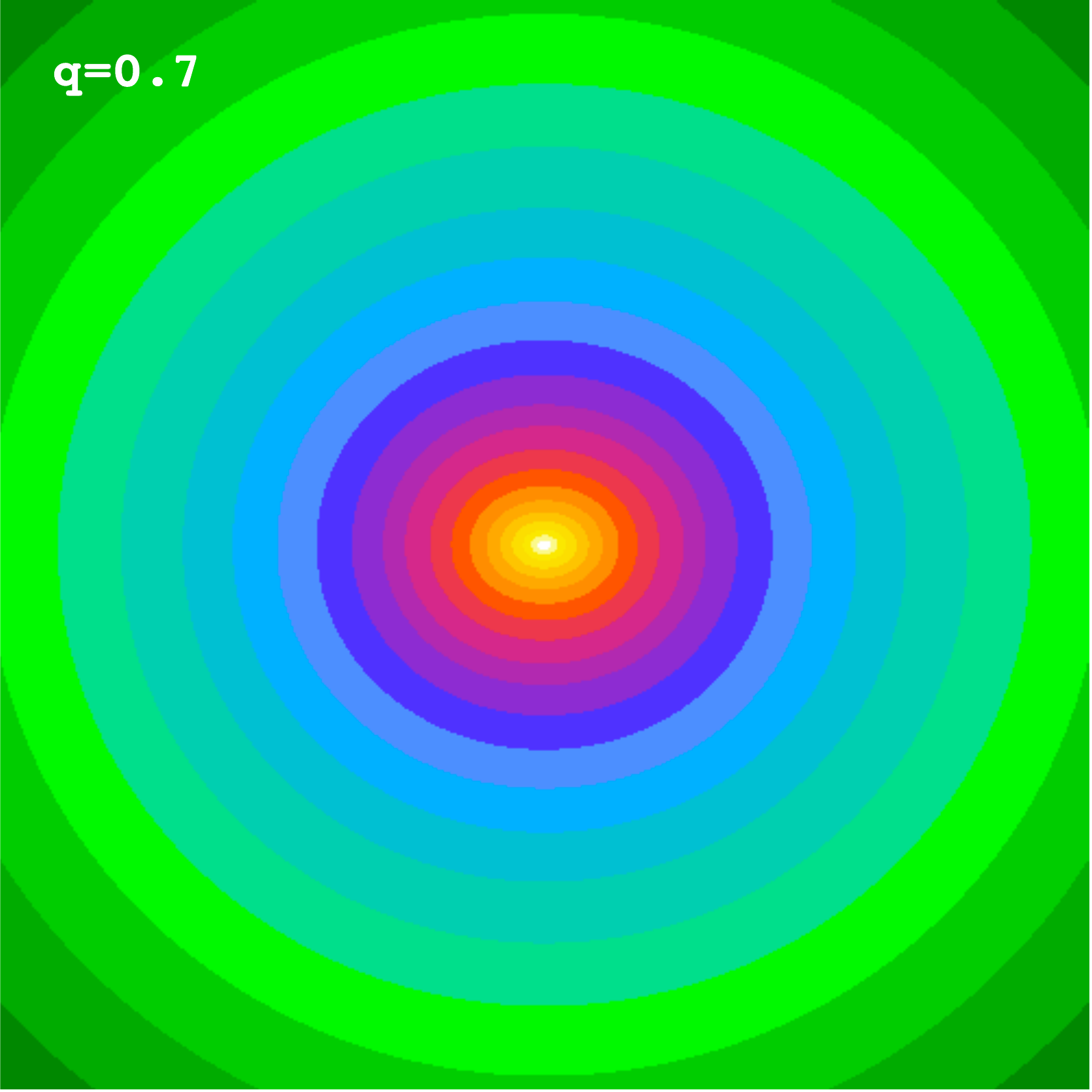}
  \\[0.2mm]
  \includegraphics[width=0.245\textwidth]{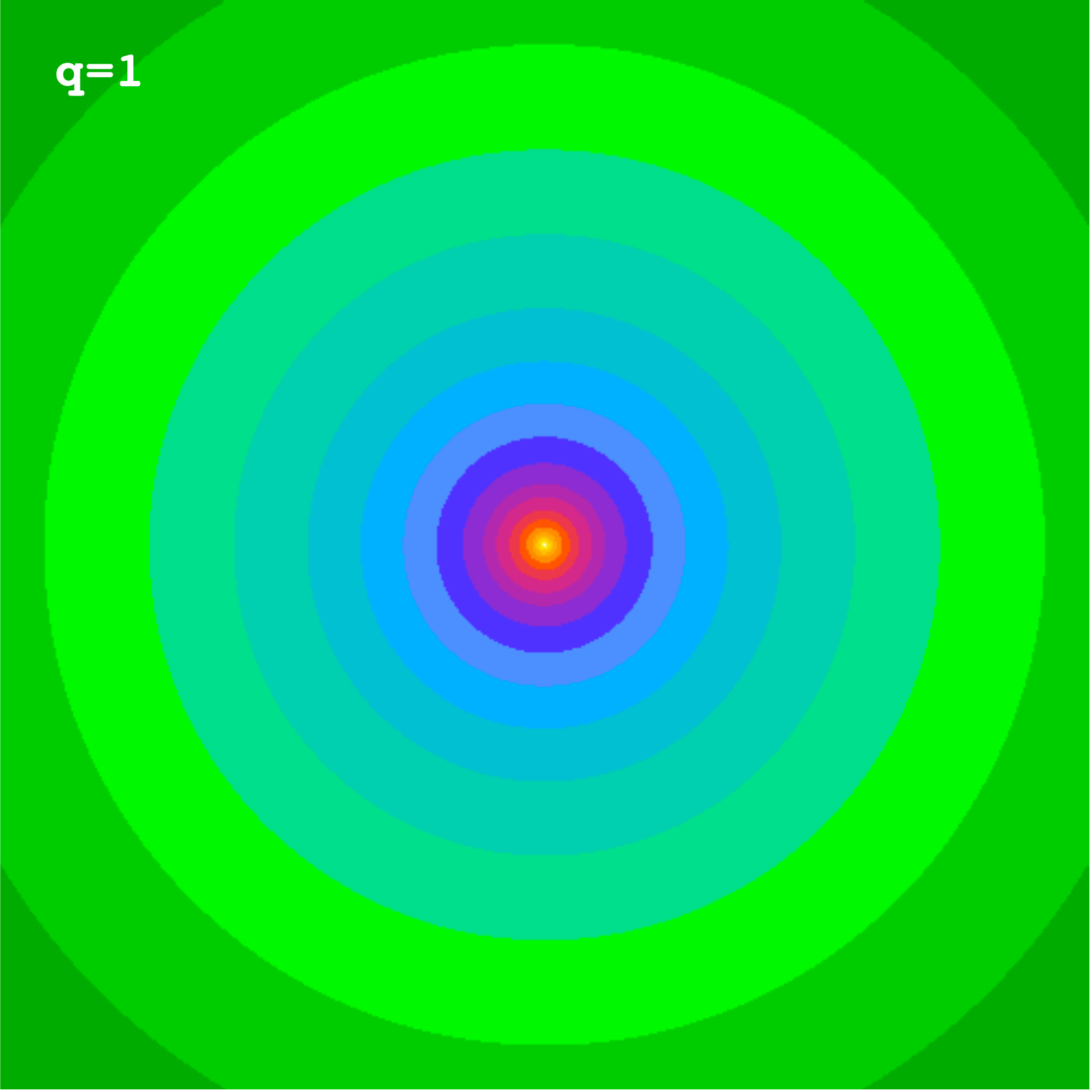}
  \includegraphics[width=0.245\textwidth]{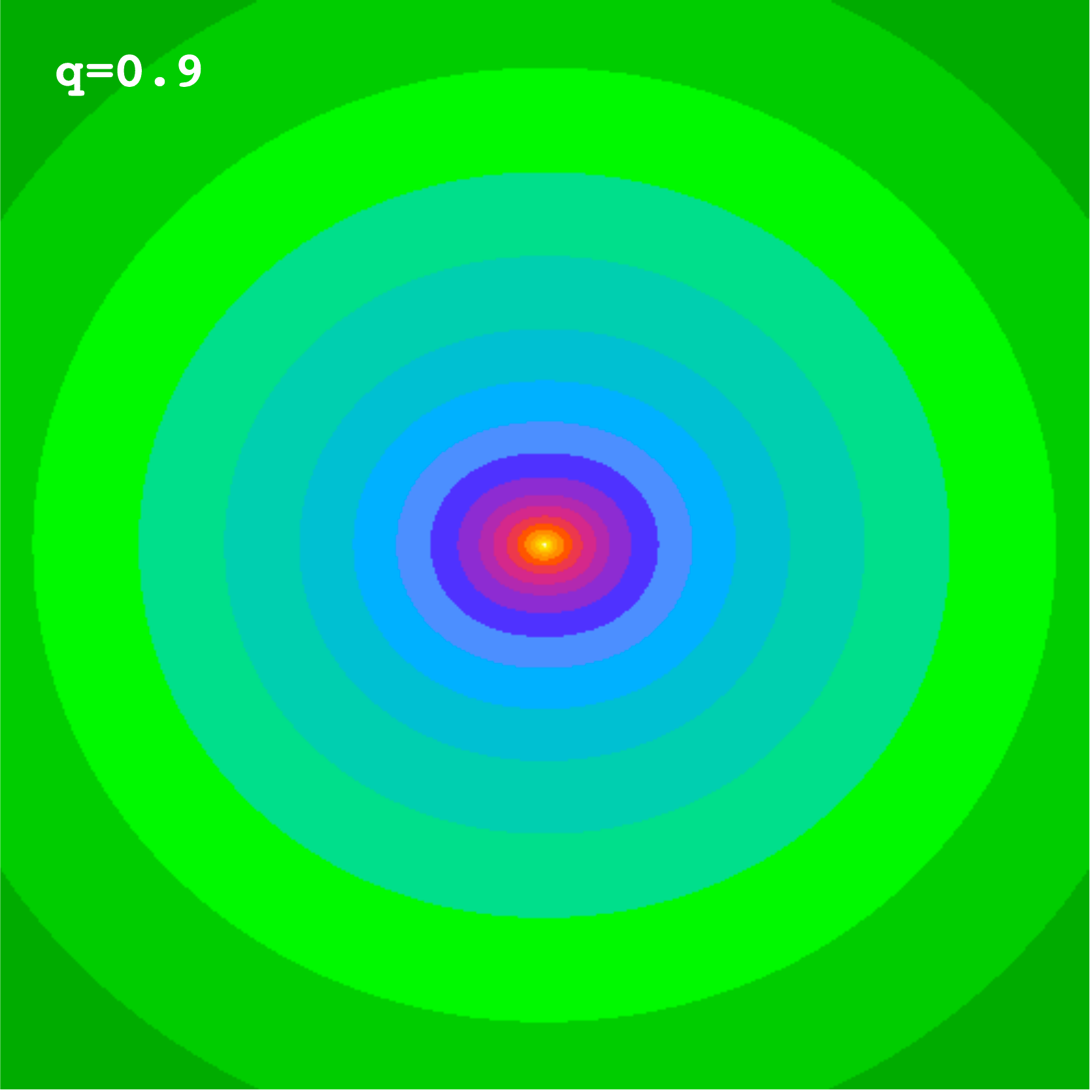}
  \includegraphics[width=0.245\textwidth]{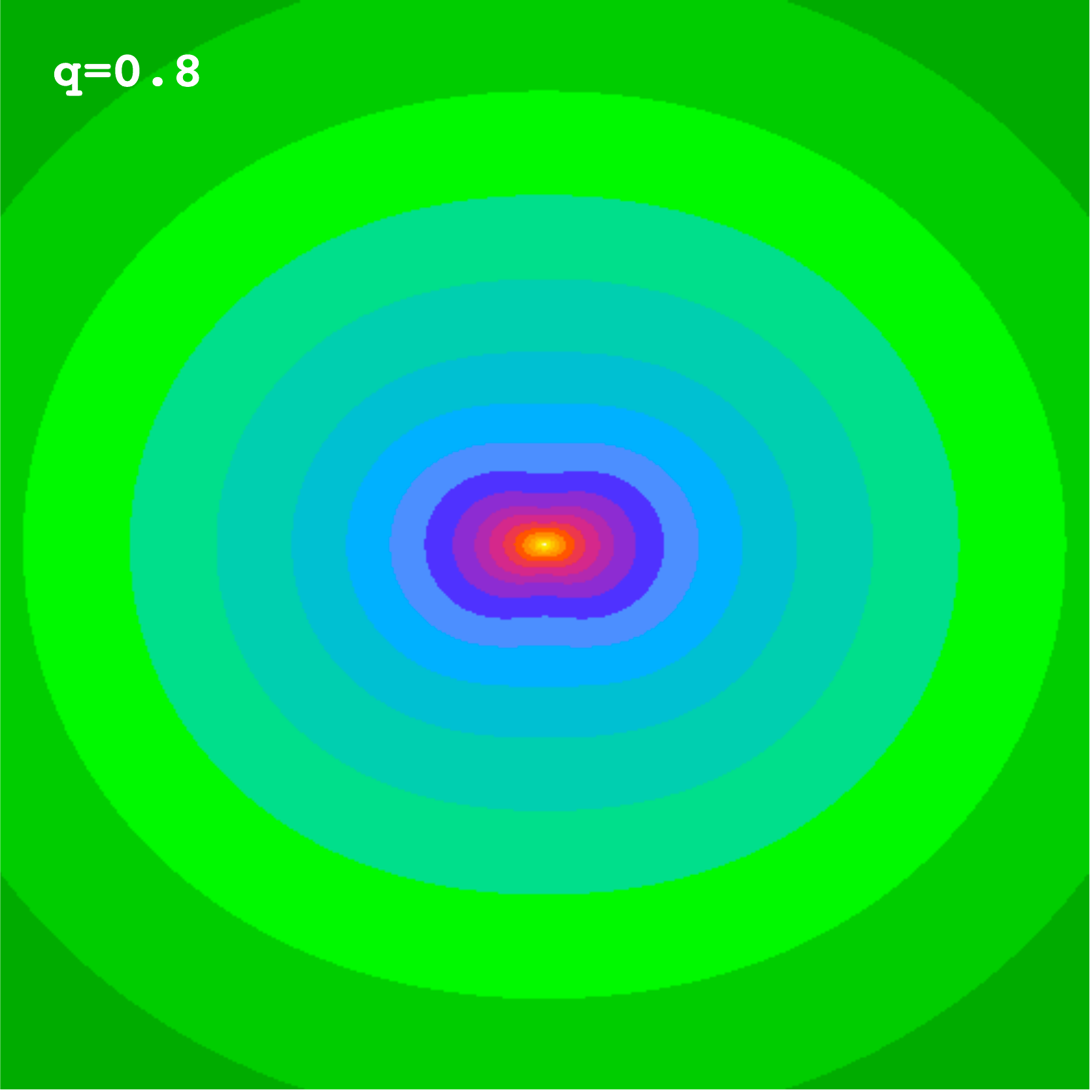}
  \includegraphics[width=0.245\textwidth]{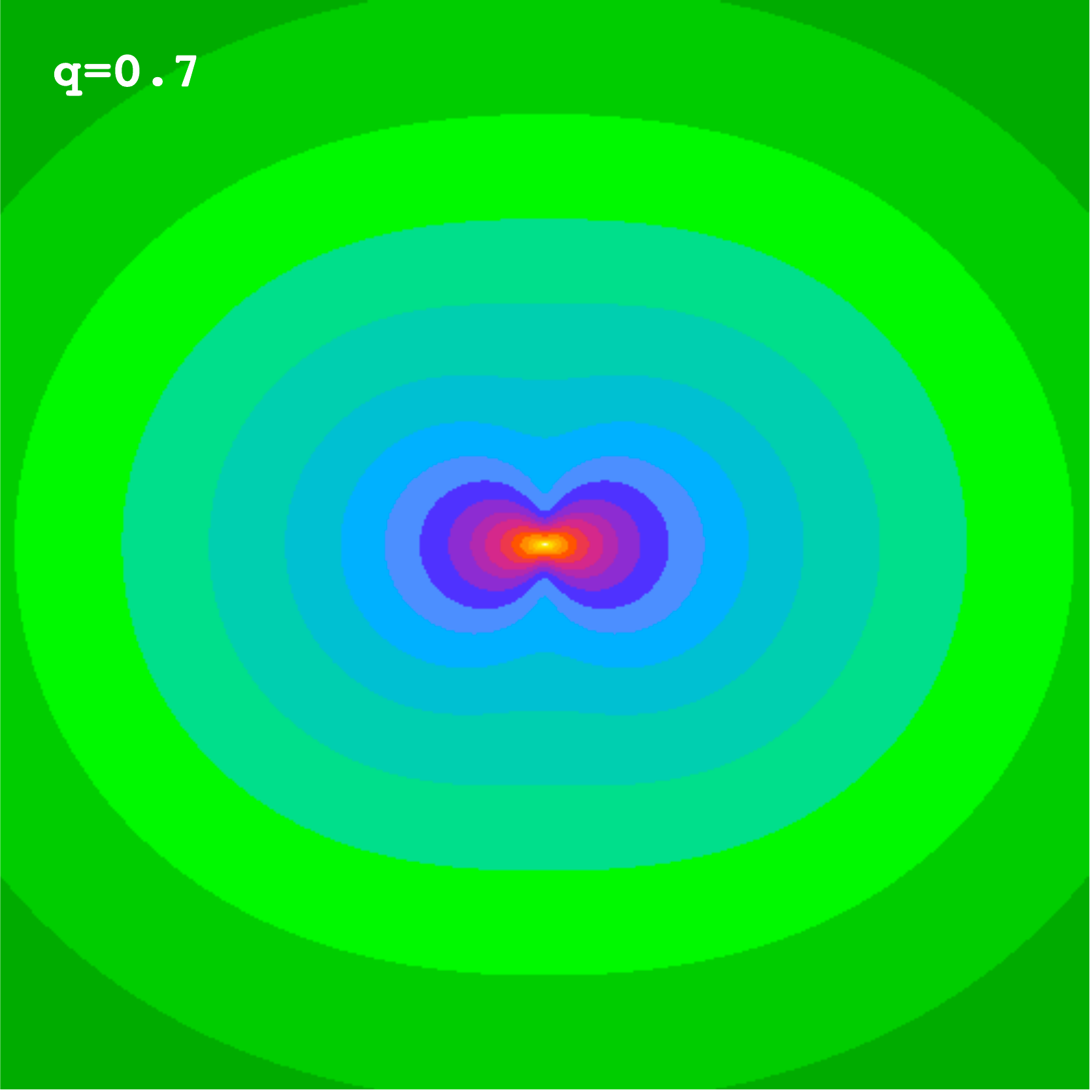}
  \caption{Similar to Figure~{\ref{potdensMN.pdf}}, but now the models
    are flattened according to the adapted recipe from
    Section~{\ref{new.sec}}. The shape of the isopotential surfaces is
    slightly smoother near the equatorial plane, which causes the
    discy structure at large radii in the isodensity surfaces to
    disappear. Instead, the isodensity surfaces remain roughly
    spheroidal with a flattening that smoothly disappears for
    increasing radii.}
  \label{potdens.pdf}
\end{figure*}

In Figure~{\ref{potdens.pdf}} we plot in a similar way the
isopotential and isodensity surfaces for flattened haloes according to
the recipe of Section~3. Compared to Figure~2, the shape of the
isopotential surfaces is hardly different, with just a small
alteration (a smoother behaviour) near the equatorial plane. This
minor change does affect the large-scale behaviour of the isodensity
surfaces significantly: the discy structure at large radii disappears
and the isodensity surfaces remain roughly spheroidal with a
flattening that smoothly disappears for increasing radii.

For models with a modest potential flattening ($0.9\lesssim q\leq1$),
all isodensity surfaces are nearly spheroidal, with a flattening
parameter $q_\rho$ ranging from roughly 0.8 at the centre to 1 at
large radii. For stronger flattening parameters, the shape of the
isodensity surfaces at small radii becomes increasingly more
peanut-shaped, as progressively more mass has be located near the
equatorial plane (and progressively less mass near the symmetry axis)
in order to stretch the isopotential surfaces into an oblate
shape. This net transfer of mass from the symmetry axis to the
equatorial plane cannot continue for ever. At a certain stage, a
negative density will arise around on either side of the equatorial
plane, causing physically unacceptable models.

\section{Discussion and conclusion}

In this paper, we have investigated methods to convert spherical
potential-density pairs into axisymmetric ones, in which the basic
characteristics of the density profile (such as the slope at small and
large radii) are retained. We attempted this by replacing the spherical
radius $r$ by an oblate radius $m$ in the expression of the
gravitational potential $\Phi(r)$. The advantage of this approach is
that the calculation of the corresponding density via Poisson's
equation requires only differentiations, such that one always obtains
fully analytical potential-density pairs. Disadvantages are that one
cannot a priori set the shape of the isodensity surfaces and that the
standard recipe for the oblate radius $m$ cannot be used since the
isopotential surfaces need to become spherical at large radii.

In Section~2 we have considered a recipe inspired by the flattening of
the Plummer potential-density pair by \citet{1975PASJ...27..533M}. We
have extended and formalized this mechanism to be applicable to
arbitrary potential-density pairs. The recipe is sufficiently simple:
the density can formally be written as a simple linear combination of
the original spherical density $\rho(m)$ and the mean density
$\bar\rho(m)$ --both evaluated at the oblate radius $m$-- and the
coefficients are simple, non-divergent functions independent of the
potential-density pair. Unfortunately, an asymptotic study
demonstrates that, at large radii, such models always show a $R^{-3}$
disc superposed on a smooth roughly spherical density distribution. As
a result, this recipe cannot be used to construct simple flattened
potential-density pairs for dynamical systems such as early-type
galaxies or dark matter haloes.

In Section~3 we have applied a modification of our original recipe
that cures the problem of the discy behaviour. The new models can be
constructed in a similar manner as the Miyamoto-Nagai type models from
Section~2; only the coefficients in the linear combination of
$\rho(m)$ and $\bar\rho(m)$ are different. An asymptotic analysis now
shows that the density distribution has the desired asymptotic
behaviour at large radii (if the density falls less rapidly than
$r^{-4}$). We also show that the flattening procedure does not alter
the shape of the density distribution at small radii: while the inner
density surfaces are flattened, the slope of the density profile is
unaltered. We have applied this recipe to construct a set of flattened
dark matter haloes based on the realistic spherical halo models by
\citet{2005MNRAS.363.1057D}. This example illustrates that the method
works fine for modest flattening values ($0.9\lesssim q\leq1$ with $q$
the flattening of the isopotential surfaces at small radii), whereas
stronger flattening values lead to peanut-shaped density
distributions.

For the sake of simplicity, we have focused this work on the
construction of flattened potential-density pairs, but there is no
reason why the recipe applied here should be limited to an oblate
geometry. The procedure is readily applicable to prolate or triaxial
geometries, which seems necessary to represent the general population
of dark matter haloes.

\end{document}